\documentclass[cite,varenna]{cimento}
\usepackage{graphicx}  
\usepackage{amsmath}
\usepackage{amsfonts}
\usepackage{amssymb}

\title{Tests of general relativity in the solar system}
\shorttitle{Tests of general relativity in the solar system}

\author{S. Reynaud}

\institute{Laboratoire Kastler Brossel, ENS, UPMC, CNRS, \\ 
Université Pierre et Marie Curie, Paris 75252 FRANCE}

\author{M.-T. Jaekel}

\institute{Laboratoire de Physique Théorique, ENS, UPMC, CNRS, \\ 
Ecole Normale Supérieure, Paris 75231 FRANCE}

\shortauthor{S. Reynaud \atque M.-T. Jaekel}

\def\beq{\begin{eqnarray}}
\def\eeq{\end{eqnarray}}
\def\beqn{\begin{equation}}
\def\eeqn{\end{equation}}

\def\stand#1{\left[#1\right]_\mathrm{GR}}

\def\s{\mathrm{s}}     % space or second
\def\d{\mathrm{d}}     % down or differential `d'
\def\u{\mathrm{u}}     % up
\def\m{\mathrm{m}}     % meter
\begin{document}

\maketitle

\begin{abstract}
Tests of gravity performed in the solar system show a good agreement 
with general relativity. 
The latter is however challenged by observations at larger,
galactic and cosmic, scales which are presently cured by introducing 
``dark matter'' or ``dark energy''.
A few measurements in the solar system, particularly the so-called
``Pioneer anomaly'', might also be pointing at a modification of gravity law 
at ranges of the order of the size of the solar system. 

The present lecture notes discuss the current status of tests of 
general relativity in the solar system.
They describe metric extensions of general relativity which 
have the capability to preserve compatibility with existing gravity 
tests while opening free space for new phenomena. 
They present arguments for new mission designs and new space technologies 
as well as for having a new look on data of existing or future experiments.
\end{abstract}

\section*{Introduction}

These notes correspond to the first lecture in a series of three~\cite{Varenna2}
given during the \textsc{International School of Physics ``Enrico Fermi''} on 
\textsc{Atom Optics and Space Physics} held at Varenna in July 2007.

They present experimental tests of gravity performed in the solar system 
which show a good agreement with general relativity (GR). 
To this aim, they first recall the basic features of the theoretical description 
and briefly review the experimental evidences supporting it as well as 
the challenges standing ahead of it.
They then discuss some perspectives in terms of theoretical ideas,
new mission designs or technological developments and
data analysis for existing or future experiments.

These notes cannot cover completely
general relativity and its impressive applications in various domains,
which have been the topic of a number of books or reviews
(see for example~\cite{Eddington,Landau,Weinberg,Misner,Will,Damour,EhlersVarenna}).
They are focused on the discussion of topics selected for
their direct connection with the question of gravity tests performed 
with tools such as atomic clocks, accelerometers, radioscience or 
laser links (see for example%
~\cite{Pound99,Shapiro,Schleich,Tourrenc,Lammerzahl01,Dittus06,BordeVarenna,%
LammerzahlVarenna,whitepaper2007}).

\section{The basic ideas in general relativity}

General relativity is built up on two basic ideas which should not be confused.
The first one is the metrical (geometrical) interpretation of gravitation, 
which was proposed soon after 1905~\cite{EquivPrinc}. 
This identification of gravitation field with the space-time metric is the 
very core of GR, but it is not sufficient to fix the latter theory. 
In order to select GR out of the variety of metric theories of gravitation, 
it is necessary to fix the relation between the geometry of space-time and its 
matter content. In GR, this relation is given by the Einstein-Hilbert equation 
which was written in 1915~\cite{EinsteinHilbert}.   

The geometrical interpretation of gravity, often presented under
the generic name of the \textit{equivalence principle}, is one of the most 
accurately verified properties of nature~\cite{Will}.
Freely falling test masses follow the geodesics of the Riemannian space-time,
that is also the curves which extremize the integral 
\beq
\int \d s \quad,\quad \d s^2\equiv g_{\mu\nu}\d x^\mu \d x^\nu 
\eeq
$g_{\mu\nu}$ is the metric tensor characterizing the space-time 
and $\d x^\mu$ the displacements in this space-time. 
As free motions obey a geometrical definition, 
they are independent of the compositions of the test masses.
This ``universality of free fall'' property has been verified with
an extreme accuracy. Its potential violations 
are parametrized by a relative difference $\eta$ 
in the accelerations $a_1$ and $a_2$ undergone by two test bodies 
in free fall from the same location with the same velocity. 
Modern experiments constrain the parameter $\eta$ to stay below 
$10^{-12}$, this accuracy being attained in laboratory 
experiments~\cite{Adelberger03} as well as in space tests using lunar laser 
ranging~\cite{LLR} or planetary probe tracking~\cite{tracking}.
These results do not preclude the possibility of violations of the equivalence
principle and such violations are indeed predicted by unification 
models~\cite{violationsEP}. 
Large improvements of this accuracy are expected in the future thanks to 
the existence of space projects \textsc{Microscope}~\cite{Microscope}
or, on a longer term, \textsc{Step}~\cite{Step}.

As another consequence of the geometrical interpretation of gravity,
ideal atomic clocks working on different transitions measure the same time,
because it is also a geometrical quantity, namely the 
proper time $\int \d s$ along the trajectory~\cite{TFmetrology,TFrelativity}.
This ``universality of clock rates'' property has also been verified with
an extreme accuracy.  Its potential variations are measured as 
a constancy of relative frequency ratios between different clocks
at a level of the order of a few $10^{-16}$ per year~\cite{varconstants}. 
These results can also be interpreted in terms of a potential ``variation
of fundamental constants''~\cite{Flambaum}. 
In this domain also, large improvements of the accuracy can be expected 
in the future thanks to the existence of the space project \textsc{Pharao-Aces}%
~\cite{Salomon} and, on a longer term, of ambitious projects 
using optical clocks~\cite{Schmidt-Kaler} or atomic sensors%
~\cite{Ertmer,Kasevich,Landragin}.

As already stated, GR is selected out of the large family 
of metric theories of gravity by the Einstein-Hilbert equation which
fixes the coupling between curvature on one hand, matter on the other one.
In order to write this equation, let us recall that there exists one tensor
in Riemannian geometry, the Einstein curvature tensor $G _{\mu\nu}$,
which has a null covariant divergence 
\beq
\label{bianchi}
&&G _{\mu\nu} \equiv R _{\mu\nu} - {1\over2} g _{\mu\nu} R\quad,\quad
D^\mu G_{\mu\nu}\equiv0 
\eeq
($R _{\mu\nu}$ is the Ricci tensor and $R$ the scalar curvature).
This geometrical identity has to be compared with the physical property 
which expresses conservation of energy and momentum as the condition of null 
divergence of the stress tensor $T _{\mu\nu}$
\beq
\label{conservation_law}
&&D^\mu T_{\mu\nu}\equiv0 
\eeq
This relation is a necessary and sufficient condition 
for free motions to follow geodesics.
The Einstein-Hilbert equation sets $G _{\mu\nu}$ and $T _{\mu\nu}$
to be proportional to each other 
\beq
\label{GR_gravitation_law}
&&G_{\mu\nu} = {8\pi G  \over c^4} T _{\mu\nu}
\eeq
The coefficient is fixed by the Newtonian limit and 
determined by the Newton constant $G$ and the velocity of light $c$.

Though relation (\ref{GR_gravitation_law}) can be used to deduce (\ref{conservation_law}) 
from the identity (\ref{bianchi}), it is not possible to deduce  
its specific form only from the geometrical interpretation of gravity.
In other words, there exists a variety of metric theories of gravity 
and GR has to be selected out of it by comparing predictions drawn from
the Einstein-Hilbert equation to the results of observations or experiments. 
When performed in the solar system, the tests effectively show a good agreement with GR,
which means that the metric tensor has a form close to its prediction. 
This statement is made more specific in the next sections.

\section{Post-Newtonian gravity tests in the solar system}

The most common approach to gravity tests in the solar system
is the so-called ``parametrized post-Newtonian'' (PPN) approach
introduced by Eddington~\cite{Eddington} and then developed by  
a number of physicists~\cite{PPN}.

The main idea can be described by coming back to the 
GR solution in the solar system.
For simplicity, we write this solution with a simple model
of the solar system where the gravity sources are reduced to the Sun 
treated as a point-like motion-less mass $M$,
and we use the Eddington gauge convention 
\beq
\label{isotropic}
&&d\s^2 = g _{00}  c ^2 d t ^2 +  g _{rr} \left( dr^2  +
 r ^2(d\theta^2 + {\rm \sin}^2\theta  d\varphi^2) \right)
\eeq
Equation (\ref{isotropic}) is the general form of an isotropic and stationary
metric, written with a specific gauge convention where spatial coordinates
are isotropic~; spherical coordinates are used with the time $t$, 
the radius $r$ (in this specific gauge), the colatitude and azimuth angles
$\theta$ and $\varphi$~; other gauge choices, such as the one leading to
the Schwartzschild solution~\cite{Schwarzschild}, would lead to the same results 
for the physical observables. 

With these simplifications, an exact solution can be found for the 
Einstein-Hilbert equation (\ref{GR_gravitation_law}) in our simple model of
the solar system.
It may be expressed through rational forms of the reduced Newton potential $\phi$
\beq
\label{GR_metric}
&&g _{00} = \left(\frac{1+\phi/2}{1-\phi/2}\right)^2\;, \quad
 g _{rr} = - \left(1-\phi/2\right)^4\;, \quad
\phi \equiv -{GM\over rc^2} 
\eeq
Note that the dimensionless $\phi$ is much smaller than unity
everywhere in the solar system. The length scale $GM/c^2$ has indeed
a value of the order of 1.5km (with $M$ the mass of the Sun), so that
the value of $\phi$ created by Sun on Earth is close to $1\times10^{-8}$.
It is therefore convenient to rewrite (\ref{GR_metric}) 
as a power series expansion in the small $\phi$
\beq
\label{GR_metric_expansion}
&&g _{00} = 1 + 2 \phi + 2 \phi^2 + \ldots \,,\quad 
 g _{rr} = -1 + 2 \phi + \ldots 
\eeq

The family of PPN metrics can then be introduced
by inserting constants in front of the terms of the power expansion
\beq
\label{PPNfamily}
&& g _{00} = 1 + 2 \alpha \phi + 2 \beta \phi^2 + \ldots \,,\quad 
 g _{rr} = -1 + 2 \gamma \phi + \ldots 
\eeq
The first parameter $\alpha$ can be set to unity by fixing Newton constant $G$
to its effective value in the solar system.
The two other parameters $\beta$ and $\gamma$ then characterize various
members of the PPN family, with GR corresponding to $\beta=\gamma=1$.
The values of these Eddington parameters affect the predicted motions, 
\textit{i.e.} the geodesics associated with the metric (\ref{PPNfamily}),
and can therefore be confronted to the observations. 
The results are conveniently presented in terms of likelyhood
plots for the anomalies $\gamma-1$ and $\beta-1$ which show
that tests favor GR as the description of gravity in the solar system. 

Doppler ranging on probes in the vicinity of Mars~\cite{tracking} and 
deflection measurements using VLBI astrometry~\cite{vlbi} have led to 
more and more stringent bounds on $\vert\gamma-1\vert$.
The current bound is essentially given by the experiment performed
through radar ranging of the Cassini probe during its 2002 solar 
occultation~\cite{cassini} 
\beq
\label{gamma}
&&\gamma - 1 = \left( 2.1 \pm 2.3\right) \times 10^{-5}
\eeq
In the meantime, analysis of the precession of planet perihelions~\cite{Talmadge88} 
and of lunar laser ranging data~\cite{LLR04}  
have led to determinations of linear superpositions of $\beta$ and $\gamma$, 
resulting in bounds on the anomaly $\vert\beta-1\vert$.
The current bound is essentially given by lunar laser ranging
\beq
&&4\left(\beta -1\right)-\left(\gamma-1\right) = \left( 4.4 \pm 4.5\right) \times 10^{-4} 
\eeq
When using (\ref{gamma}), this leads to~\cite{LLR04} 
\beq
\label{beta}
&& \beta - 1 = \left( 1.2 \pm 1.1\right) \times 10^{-4}
\eeq

Equations (\ref{gamma},\ref{beta}) constitute a quantitative statement
of the agreement with GR of gravity tests in the solar system.
Let us emphasize that their common presentation under the form
``general relativity is confirmed by the tests'' is a bit too loose.
The tests discussed so far do not answer by a final ``yes'' answer
to a mere ``yes/no'' question. 
They rather select a vicinity of GR, defined by (\ref{gamma},\ref{beta}),
as the best current description of gravity within the family of PPN metrics. 
This warning is not a mere precaution, it is rather a pointer to
possible future progress in the domain.
There indeed exist theoretical models~\cite{violationsGamma} which 
deviate from GR while staying within the bounds (\ref{gamma},\ref{beta}).
Furthermore, as discussed below, extensions of GR do not necessarily belong
to the PPN family (\ref{PPNfamily}).

This last point is in particular emphasized by the so-called ``fifth force'' tests 
which are focused on a possible scale-dependent deviation from the gravity
force law. Such deviations, predicted by some unification models, remain however 
unobserved to date~\cite{Fischbach98}. 

\section{Scale-dependent tests in the solar system}

The main idea is here to check the $r-$dependence of the gravity  
potential, that is also of the component $g _{00}$.
Hypothetical modifications of its standard expression, predicted by unification models,
are often parametrized in terms of an additional Yukawa potential 
\beq
\label{Yukawa}
&& g _{00} = \stand{g _{00}} + \delta g _{00} \,,\quad
\delta g _{00} = 2\phi(r) \ \alpha \exp\left(-\frac r\lambda\right)
\eeq
This potential depends on two parameters, an amplitude $\alpha$ measured with respect to 
Newton potential and a range $\lambda$ related through a Yukawa-like relation to
the mass scale of the hypothetical new particle which would mediate the ``fifth force''.

The presence of such a correction has been looked for on a large range of distances.
When the whole set of results is reported on a global diagram (printed as Fig.~1 in
Ref.~\cite{JR04} where it was reproduced thanks to a courtesy of Coy \etal \cite{Coy03}),
it appears that the Yukawa term is excluded with a high accuracy at some ranges 
\beq
\label{bestYukawa}
\alpha<10^{-10}\; &\mathrm{at}& \;\lambda\simeq\mathrm{Earth-Moon\ distance} \\
\alpha<10^{-9}\; &\mathrm{at}& \;\lambda\simeq\mathrm{Sun-Mars\ distance} 
\eeq
These bounds, again deduced from lunar laser ranging~\cite{LLR}
and tracking of planetary probes~\cite{tracking}, correspond to a remarkable 
result which approaches the accuracy of equivalence principle tests.
It is also clear that windows remain open for large Yukawa corrections ($\alpha>1$)
at short ranges as well as long ranges~\cite{JR04}.

These two windows are being actively explored.
The accuracy of short range tests has been recently improved, as gravity experiments 
were pushed to smaller and smaller distances~\cite{shortrange}.
The strongest constraints at sub-millimeter ranges come from experiments
with torsion pendula and have recently reached an impressive 
performance~\cite{Kapner07}
\beq
\label{shortYukawa}
&& \alpha < 1\;\mathrm{for}\;\lambda > 56\mu\m 
\eeq
At even shorter ranges, tests of the gravity force law are pursued as careful
comparaisons between theoretical predictions and experimental measurements 
of the Casimir force, which becomes dominant at micrometric ranges~\cite{casimir}.

In the long range window, a test of the gravity force was initiated by NASA 
as the extension of Pioneer 10/11 missions and it led to the observation 
of the Pioneer anomaly, the most notable exception of a signal in conflict
with the predictions of GR.
Before embarking in the discussion of this signal (in the next section),
let us emphasize that the apparent contradiction between Pioneer observations
and other gravity tests may be cured in an extended framework, where
deviations from GR may show a scale dependence.
This corresponds to a situation where the accurate tests on 
the Sun-Earth-Moon and Sun-Mars systems would reproduce GR predictions 
whereas tests at different scales could show deviations.

In fact, this idea has to be considered with great attention in the 
current context of fundamental physics where questions naturally arise
about the validity of GR at galactic or cosmic scales.
It is well known that the rotation of the outer regions of galaxies
cannot be reproduced by using GR for gravity and fixing the matter content 
as we see it from other observation channels.
This gravitational anomaly is commonly accounted for by introducing 
``dark matter'' tuned to reproduce the rotation curves~\cite{darkmatter}. 
{}A further anomaly has been detected more recently as an acceleration of 
cosmic expansion and it is interpreted as due to the presence of some 
``dark energy'' component~\cite{darkenergy}.
Both components of the ``dark side'' of the universe have no universally 
accepted interpretation nor are they observed through other means 
than the gravitational anomalies they have been designed to cure%
\cite{Wetterich}. 
As long as this situation is lasting, these anomalies may as well be interpreted 
as long range deviations from GR~\cite{longrangemodifGR}.

It is therefore important to make any possible effort to test
gravity at the largest scales attainable by man made instruments, in an attempt
to bridge the gap between gravity experiments in the solar system and
observations at the much larger galactic and cosmic scales. 

\section{The Pioneer anomaly}

The best example of such a strategy to date was the NASA decision 
to extend Pioneer 10 \& 11 missions after their primary planetary objectives
had been met~\cite{Fimmel}.
The idea was from the beginning to use the two precisely navigated deep space
vehicles to test the laws of gravity at large heliocentric distances and
to search for low-frequency gravitational waves.
The latter objective failed to detect gravitational waves~\cite{Armstrong} 
but the former one led to the largest scaled test of gravity ever carried out.
The most notable output of this experiment is certainly that it failed to 
confirm the known laws of gravity~\cite{Anderson98,Anderson02}.

The anomaly was recorded on tracking data~\cite{Asmar05} 
of the Pioneer 10 \& 11 probes by the NASA deep space network (DSN).
An up-link radio signal is emitted from Earth at a DSN station, it is then 
transponded back by the probe, and the down-link radio signal is finally 
received on Earth at the same or another DSN station.
{}For Pioneer 10 \& 11 probes, the observable was the Doppler shift, 
defined as the ratio of cycle counting rates at the down and up events,
as measured by reference clocks located at reception and emission stations.
This observable can equivalently be interpreted as a Doppler velocity 
$\upsilon$ measuring a relative velocity of the probe with respect to the 
station
\beq
\label{DopplerVelocity}
&&\frac{f_\d}{f_\u}\equiv \frac{1-{\upsilon\over c}}{1+{\upsilon\over c}}
\eeq
Note that this is just an interpretation of the observable, which contains
not only the effect of motion but also relativistic and gravitational effects. 
It is also worth keeping in mind that perturbations due to transmission 
media effects have to be accounted for~\cite{Moyer03}.
  
These Doppler tracking data were analyzed during the cruises of Pioneer 10 \& 11.
When the probes had flown by the giant planets
(Jupiter for Pioneer 10, Jupiter and Saturn for Pioneer 11) 
and reached a quieter environment, 
it was noticed that tracking data were deviating from the predictions of GR.
Precisely, the observed Doppler velocity was departing from the modelled 
Doppler velocity (see the Fig.~8 of Ref.~\cite{Anderson02}).
As the deviation $\delta\upsilon$ was varying linearly with elapsed time,
it was natural to introduce an anomalous acceleration to describe it
\beq
\label{anomalyP}
&&\delta\upsilon \equiv \upsilon_\mathrm{observed} - \upsilon_\mathrm{modelled} 
\simeq -a_P \left( t - t_\mathrm{in} \right) 
\eeq
Note that the anomalous acceleration $a_P$ does not necessarily reflect
a force acting on the probe. But, when considered as such, it appears
to be directed towards the Sun with an approximately constant amplitude 
over a large range of heliocentric distances (AU $\equiv$ astronomical unit)
\beq
\label{Pioneer_acceleration}
&&a_P = (0.87 \pm 0.13) ~\mathrm{nm}~\mathrm{s}^{-2}\quad ,\quad 
20~\mathrm{AU}\lesssim r_P \lesssim 70~\mathrm{AU}  
\eeq

The Pioneer anomaly has been registered on the 
two deep space probes showing the best navigation accuracy.
Anomalous observations have also been reported for Ulysses and Galileo
probes, but not with the same reliability as for Pioneer probes.
{}For other probes like Voyager 1 \& 2, the navigation 
accuracy was not sufficient.
This means that the Pioneer gravity test has been performed twice with
identical probes on similar trajectories - escape directions opposite
in the solar system - and the same result - similar Sunward acceleration.
This is not an impressive statistics when compared to the large number
of tests confirming GR.
In particular, when the possibility of an artefact onboard the probe 
is considered, this artefact could be the same on the two probes.

The extensive analysis of the JPL team has been published after years of cross 
checks~\cite{Anderson02} and the presence of the anomaly has been confirmed by 
several independent analysis~\cite{PAconfirmations}.
A number of mechanisms have been considered as attempts of explanations of the
anomaly as a systematic effect generated  by the spacecraft itself or its 
environment (\cite{PAexplanations} and references therein) but they have not
led to a satisfactory understanding to date.
Further informations are expected to come out of the re-analysis process now going on
with recently recovered Pioneer data~\cite{PAdatarecovery,ISSI} (more details below).

The Pioneer anomaly constitutes an intriguing piece of information in a 
context where the status of gravity theory
is challenged by the puzzles of dark matter and dark energy.
If confirmed, this signal might already reveal an anomalous behaviour of gravity 
at scales of the order of the size of the solar system and thus have a 
strong impact on fundamental physics, astrophysics and cosmology.
It is therefore important to investigate what could be the signatures of 
the Pioneer anomaly as a long range modification of gravity
(see for example~\cite{JaekelReynaud,JRcqg06,Moffat,Bertolami,Lammerzahl} 
and references therein).

\section{Can the Pioneer signal be a metric anomaly ?}

If there exist gravity theories where a Pioneer signal can take a natural place, 
they must be considered with great care.
If on the contrary one can prove that there exist no such 
theories, this result is also important since it allows the
range of validity of GR to be extended to the whole solar system. 
Anyway, the relevance of the Pioneer anomaly for space navigation is already
sufficient to deserve a close scrutiny.

As we know that tests of the equivalence principle (EP) have shown it to be 
preserved at the level of a few 10$^{-13}$ in lunar laser ranging, 
it appears quite unlikely that an EP violation could 
account for the Pioneer anomaly.
To make this statement more precise, let us say that the standard Newton 
acceleration is of the order of 1~$\mu$m~s$^{-2}$ at 70~UA 
so that the Pioneer anomaly is only one thousand times smaller.
This remark does not contradict the possibility of EP violations which are 
predicted by unification models and looked for in space experiments such as 
MICROSCOPE and STEP, but however suggests that such violations are expected 
to occur at a too small level to produce the Pioneer anomaly.
In the following, we therefore restrict our attention 
to the confrontation of GR with alternative metric theories of gravity.

In this well understood metric framework, 
the form of the coupling between curvature and matter 
can still be discussed~\cite{JaekelReynaud}.
Like the other fundamental interactions, gravitation may be treated
as a field theory~\cite{GRfieldtheory} and
radiative corrections due to its coupling to other fields thus naturally lead 
to consider GR within a larger class of theories~\cite{GRradiativecorrections}.
Modifications are thus expected to appear in particular 
at large length scales~\cite{GRlongrangemodifs}.
Note that the embedding of GR in the larger family of fourth order theories 
cures its unrenormalizability while providing asymptotic freedom%
~\cite{GRrenormalizability}.
This is a strong argument for extending the gravitation theory at scales 
not constrained by experiments, for instance 
using renormalization group trajectories~\cite{Reuter}.
Renormalizability of these theories however comes with a counterpart, that is the 
problem of ghosts. 
It has however been argued that this problem does not constitute a deadend for an effective 
field theory and, in particular, that the departure from unitarity is expected to 
be negligible at ordinary scales tested in present day universe~\cite{GRunitarity}.

The main message at this point is that GR, as other current theories of fundamental 
interactions, has to be considered as an effective theory of gravity, 
certainly valid at the length scales for which it has been accurately tested, 
but still to be tested at other scales. 
Within the metric framework, it is possible to develop a well founded phenomenological 
point of view for analyzing the potential meaning of the Pioneer anomaly~\cite{JRcqg06}.
This framework covers the whole spectrum of metric extensions of GR 
which remain in the vicinity of GR and it includes as particular cases 
the PPN extensions as well as the ``fifth force'' extensions of GR.
It can be considered as a natural consequence of radiative corrections due to the 
coupling of gravity with other fields~\cite{GRradiativecorrections}
and has also recently been shown to be reproduced by 
field theoretical extensions of GR~\cite{BrunetonEsposito}.

\section{Post-Einsteinian metric extensions of GR}

We refer the interested readers to~\cite{JRcqg06} 
for a complete description of this framework.
Here we rapidly present some of its salient features 
and phenomenological consequences.

As we consider an isotropic and stationary model of the solar system with gravity
field created by a single point-like and motion-less source, the Sun, 
the extended metric can be written in terms of two potentials
\beq
\label{extended_metric}
&&\d s^2 = g_{00}(r) c^2 \d t^2 + g_{rr}(r) 
\left(\d r^2 + r^2 \d\theta^2 + r^2 \sin^2\theta \d\varphi^2\right) \\
&&g_{00}(r) \equiv \stand{g_{00}(r)} + \delta g_{00}(r)  \quad,\quad 
g_{rr}(r) \equiv \stand{g_{rr}(r)} + \delta g_{rr}(r)  \nonumber
\eeq
This expression has been written in the Eddington isotropic gauge (it could have been
written as well with other gauge choices). The GR metric $\stand{g_{\mu\nu}}$ may be 
treated exactly (see eq.\ref{GR_metric}), or at a sufficient order in the power 
expansion (\ref{GR_metric_expansion}).
Finally the anomalous part $\delta{g_{\mu\nu}}$  of the metric may be treated
at first order, since we consider the solution to remain in the vicinity of GR. 
The extension of GR is thus parametrized by two functions, say $\delta g_{00}(r)$ and 
$\delta \left(g_{00}g_{rr}\right)(r)$, which describe a much wider phenomenology than
the PPN framework which was parametrized by two constants. 

It is also possible to consider that the post-Einsteinian extensions 
generalize the PPN metrics by promoting the two constants $\beta,\gamma$ 
to the status of functions of $r$. 
In fact, the two functions accomodate the freedom coming with the
extension of the coupling between curvature and matter.
This extension is characterized by two running couplings which depend 
on scale and cover the two sectors of the gravity field corresponding
to traceless (conformal weight 0) and traced components (conformal weight 1)
of the metric tensor~\cite{GRradiativecorrections}. 
Equivalently, it may be characterized by the variation with $r$
of two components of the Einstein curvatures which no longer vanish 
outside the source, as they did in GR.
Anew, the PPN family is a particular case which already shows such an 
anomalous behaviour of Einstein curvatures,
with however a specific $r-$dependence.

In phenomenological terms, the function $\delta g_{00}$ 
represents an anomaly of the gravity force analogous to a ``fifth force''. 
It has to remain small enough to preserve the good agreement of 
gravity tests performed on planetary orbits with GR (see the discussion 
of ``fifth force'' tests above).
Meanwhile, the second function $\delta \left(g_{00} g_{rr}\right)$ 
represents an extension of PPN phenomenology with a $r-$dependent 
Eddington parameter $\gamma$. 
It opens an additional freedom with respect to the mere 
modification of the gravity potential $g_{00}$ as well as with respect
to the PPN metrics where $\gamma$ is constant. 
This additional phenomenological freedom makes it possible to accomodate
the Pioneer signal without raising conflicts with other gravity 
tests~\cite{JaekelReynaud}.

If one supposes for simplicity that $\delta g_{00}$ follows a Yukawa law,
it is easy to show that it cannot explain the Pioneer signal and simultaneously have
escaped detection in the constraining tests performed with martian probes~\cite{JR04}.
But $\delta g_{00}$ could appear only after Saturn 
(this is suggested by Fig.~7 in~\cite{Anderson02})
while being cut at smaller distances~\cite{Moffat}.
Some authors have argued that the ephemeris of outer planets were accurate enough 
to discard the presence of an anomaly at distances explored by the Pioneer probes
\cite{Iorio_etal}. They have pushed their claim 
one step farther by objecting to the very possibility 
of accounting for the Pioneer anomaly in any metric theory of gravity. 
This claim is clearly untenable at the moment. 
It is only after a careful comparison of observations in the outer solar system 
with the post-Einsteinian phenomenological framework that it will be possible to 
know whether or not the latter is able to fit all gravity tests,
including the Pioneer one. 

This comparison will have to account for the presence of the two 
potentials as well as for their $r-$dependences.
The results of the calculations in~\cite{JRcqg06} 
(after the correction of a mistake in previous papers~\cite{JaekelReynaud}),
show that a Pioneer-like signal can be produced either by an anomaly
proportional to $r$ in $\delta g_{00}$ or by an anomaly
proportional to $r^2$ in $\delta \left(g_{00}g_{rr}\right)$. 
A linear superposition of the two possibilities can also be considered. 
As the two possibilities produce different signatures, they can
in principle be distinguished in the comparison with tracking data.
This situation certainly pleads for pushing the study farther and 
comparing the theoretical expectations with the Pioneer data.

\section*{Conclusion}

In order to address the challenge raised by the existence of the Pioneer anomaly,
one has different strategies at one's disposal and should use all of them~:
\begin{itemize}
\item the reanalysis of the Pioneer data is an important tool with the ability
to confirm (or not) the existence of the anomaly and learn more about
its properties; this idea is today more relevant than ever thanks to the recent 
recovery of data covering the whole period of Pioneer 10 \& 11 missions from 
launch to the last data point~\cite{PAdatarecovery}; 
not only Doppler tracking data, but also telemetry data, are available
and the latter contain valuable information on the history of the probes. 

\item the Pioneer data re-analysis program is an ongoing international joint effort 
with teams at work in USA, Canada, Germany, France, Italy and Norway~\cite{ISSI};
this effort should lead to an improved control of systematics 
and produce new information of importance on several properties of the force 
- its direction, long-term variation as well as annual or diurnal modulations, 
spin dependence$\ldots$;
of particular interest would be a clear answer to the question of the onset 
of the anomaly~\cite{onset}.

\item several proposals~\cite{PAEM,ENIGMA,ZACUTO,ODYSSEY,SAGAS} have been made 
for new missions designed to study the anomaly and understand its origin;
two missions have been proposed to the Cosmic Vision selection process at ESA,
the first one as a medium size mission~\cite{ODYSSEY} using accelerometer and 
radioscience instruments upgraded from existing technology, 
the second one as a large size mission using new quantum sensors to map the 
two components of the gravity field with high accuracy~\cite{SAGAS}. 

\item these projects are proposing new technologies 
of importance for future fundamental physics and deep space exploration;
some of the proposed instruments may be flown as passengers 
on planetary missions to Mars, Jupiter or Saturne, with the aim of meeting 
some of the objectives of deep space gravity tests;
an accelerometer on-board would measure non gravitational forces of any origin and  
solve ambiguities on thermal radiation force, or drag force; 
better radio tracking techniques or laser techniques would reduce uncertainties 
and improve control systematics; 
such passenger instruments could provide new information on the anomalous behaviours 
observed on several probes at planetary flybies~\cite{flybyanomaly}.  

\item in the meantime, it would be wise to develop and validate enabling 
technologies, such as accelerometers for measuring non gravitational
forces of any origin (thermal radiation budget on the probe, drag force...), 
radio and laser techniques for ranging/tracking, 
accurate clocks for measuring the gravity potential on board.

\item without waiting for new programs to be set up,
the theoretical studies must be pushed farther and 
their predictions compared with the Pioneer data;
this study can lead to Pioneer-related signals,
produced by the same causes, but to be looked for in the data
of other experiments;
in this respect, the GAIA project~\cite{Gaia} is of particular interest,
firstly because it will produce an accurate and global mapping of 
deflection over the sky;
such a mapping could reveal the presence of an anomalous metric 
as a dependence of the parameter $\gamma$ versus the angular distance
to the Sun (the results being reduced to PPN ones when $\gamma$ is constant);
then GAIA will also accurately track the motions of small bodies 
in the solar system and thus significantly improve the coverage
of dynamical behaviours with the possibility of confirming 
or discarding the existence of metric anomalies. 

\end{itemize}

\acknowledgments
Acknowledgments for stimulating discussions are due to a number of persons
involved in the collaborations working on this topic~:
S.G. Turyshev and the \textsc{Pioneer Anomaly Investigation} team at 
the International Space Sciences Institute~\cite{ISSI}, H. Dittus 
and the \textsc{Pioneer Anomaly Explorer Mission} team~\cite{PAEM},
U. Johann and the \textsc{Enigma} team~\cite{ENIGMA},
O. Bertolami and the \textsc{Zacuto} team~\cite{ZACUTO},
B. Christophe and the \textsc{Odyssey} team~\cite{ODYSSEY},
P. Wolf and the \textsc{Sagas} team~\cite{SAGAS}, 
P. Berio, J.-M. Courty, A. Lévy and G. Metris for the french
Pioneer data analysis team.

\def\etal{\textit{et al }}
\def\ibid{\textit{ibidem }}
\def\url#1{\textrm{#1}}
\def\arxiv#1{\textrm{#1}}
\def\REVIEW#1#2#3#4{\textit{#1} \textbf{#2} {#4} ({#3})}
\def\Book#1#2#3{\textit{#1} ({#2}, {#3})}
\def\BOOK#1#2#3#4{\textit{#1} ({#2}, {#3}, {#4})}
\def\BOOKed#1#2#3#4#5{\textit{#1}, #2 ({#3}, {#4}, {#5})}
\def\Name#1#2{\textsc{#1}~#2}

\end{document}